# On differences between fractional and integer order differential equations for dynamical games

E.Ahmed[1]  A.S.Elgazzar[2] and M.I.Shehata[1]

1. Mathematics Department, Faculty of Science, Mansoura 35516, EGYPT.
2. Mathematics Department, Faculty of Education, Al-Arish, EGYPT.

Abstract:
   We argue that fractional order (FO) differential equations are more suitable to model complex adaptive systems (CAS). Hence they are applied in replicator equations for non-cooperative game. Rock-Scissors-Paper game is discussed. It is known that its integer order model does not have a stable equilibrium. Its fractional order model is shown to have a locally asymptotically stable internal solution. A FO asymmetric game is shown to have a locally asymptotically stable internal solution. This is not the case for its integer order counterpart.

**1. Fractional order equations (FE) :**

**Definition (1)** [1]: A complex adaptive system (CAS) with emergence consists of interacting adaptive agents, where the properties of the system as a whole do not exist for the individual elements (agents) and are not caused by external effects.

**Definition (2):** "An emergent property of a CAS is a property of the system as a whole which does not exist at the individual elements (agents) level".

   Typical examples are the brain, the immune system, the economy, social systems, ecology, insects swarm, etc…

   Therefore to understand a complex system one has to study the system as a whole and not to decompose it into its constituents. This totalistic approach is against the standard reductionist one, which tries to decompose any system to its constituents and hopes that by understanding the elements one can understand the whole system. Since this is quite difficult, mathematical and computer models may be helpful in studying such systems.

   Recently [2] it became apparent that fractional equations naturally represent systems with memory. To see this consider the following evolution equation

$$df(t)/dt = -\lambda^2 \int_0^t k(t-t')f(t')dt' \quad (1)$$

If the system has no memory then $k(t-t') = \delta(t-t')$ and one gets $f(t) = f_0 \exp(-\lambda^2 t)$. If the system has an ideal memory then $k(t-t') = \{1 \text{ if } t \geq t', \ 0 \text{ if } t < t'\}$ hence $f \approx f_0 \cos \lambda t$. Using Laplace transform $L[f] = \int_0^\infty f(t)\exp(-st)dt$ one gets L[f]=1 if there is no memory and 1/s if there is ideal memory hence the case of non-ideal memory is expected to be given by $L[f] = 1/s^\alpha$, $0 < \alpha < 1$. In this case equation (1) becomes a fractional order differential equation:

$$D^\alpha f(t) = \int_0^t (t-t')^{\alpha-1} f(t')dt'/\Gamma(\alpha)$$

where $\Gamma(\alpha)$ is the Gamma function. This system has the following solution

$$f(t) = f_0 E_{\alpha+1}(-\lambda^2 t^{\alpha+1}),$$

where $E_\alpha(z)$ is the Mittag-Leffler function given by

$$E_\alpha(z) = \sum_{k=0}^{\infty} z^k / \Gamma(\alpha k + 1)$$

It is direct to see that $E_1(z) = \exp(z), E_2(z) = \cos z$.

Following a similar procedure to study a random process with memory, one obtains the following fractional evolution equation

$$\partial^{\alpha+1} P(x,t)/\partial t^{\alpha+1} = \sum_n (-1)^n / n! \partial^n [K_n(x) P(x,t)]/\partial x^n, \qquad 0 < \alpha < 1$$

where P(x,t) is a measure of the probability to find a particle at time t at position x. For the case of fractional diffusion equation the results are

$$\partial^{\alpha+1} P(x,t)/\partial t^{\alpha+1} = D \partial^2 P(x,t)/\partial x^2, \ P(x,0) = \delta(x), \partial P(x,0)/\partial t = 0 \Rightarrow$$

$$P = (1/(2\sqrt{D t^\beta})) M(|x|/\sqrt{D t^\beta}; \beta), \quad \beta = (\alpha+1)/2$$

$$M(z;\beta) = \sum_{n=0}^{\infty} [(-1)^n z^n / \{n! \Gamma(-\beta n + 1 - \beta)\}]$$

For the case of no memory $\alpha = 0 \Rightarrow M(z;1/2) = \exp(-z^2/4)$.

Moreover it has been proved that fractional order systems are relevant to fractal systems and systems with power law correlations [3].

Thus fractional equations naturally represent systems with memory and fractal systems consequently fractional order equations are relevant to CAS since memory and fractals are abundant in CAS systems.

## 2. Fractional order calculus in non-cooperative games:

Game theory [4] is the study of the ways in which strategic interactions among rational players produce outcomes (profits) with respect to the preferences of the players. Each player in a game faces a choice among two or more possible strategies. A strategy is a predetermined program of play that tells the player what actions to take in response to every possible strategy other players may use.

A game is non-cooperative if the players compete with each other hence they do not form collusions.

A basic property of game theory is that one's payoff depends on the others' decisions as well as his.

The mathematical framework of the game theory was initiated by von Neumann and Morgenstern in 1944. Also they had suggested the max-min solution for zero-sum games. Stability concepts for non-cooperative games.

To quantify this concept one may use [4] the replicator equation which intuitively means that the rate of change of the fraction of players adopting strategy i is proportional to the difference between their payoff and the average payoff of the population i.e.

$$dx_i/dt = x_i[(\Pi x)_i - x \Pi x], i = 1, 2, .., n, \sum_{i=1}^{n} x_i = 1, \ 1 \geq x_i \geq 0$$

where $\Pi$ is the payoff matrix.

Now since we have shown that fractional order equations are more suitable to model CAS, we study some examples of fractional order game theory using fractional order replicator equation:

$$D^\alpha x_i(t) = x_i[(\Pi x)_i - x\Pi x], 1 \geq \alpha \geq 0. \quad (2)$$

To study the stability of the equilibrium solutions of (2) we use the result of Matignon [5] which has been generalized in [6] that the equilibrium solutions of the system (2) are locally asymptotically stable if all the eigenvalues $\lambda$ of the characteristic roots satisfy

$$|\arg(\lambda)| > \alpha\pi/2 \quad (3).$$

Obviously this condition is a relaxation of Routh-Hurwitz conditions which correspond to $\alpha = 1$. Moreover if the system is 1-dimensional then the stability condition becomes $\lambda < 0, \forall \alpha \in (0,1]$. Hence we have

Proposition (1): The stability of equilibrium solutions for all non-cooperative symmetric games with only two strategies allowed is the same $\forall \alpha \in (0,1]$.

Now we present an example where a stable equilibrium solution does exist for $1 > \alpha > 0$ but not for $\alpha = 1$. Consider the Rock-Scissors-Paper game with payoff matrix $\begin{bmatrix} 0 & 1 & -1 \\ -1 & 0 & 1 \\ 1 & -1 & 0 \end{bmatrix}$ where strategy one is rock, strategy 2 is scissors and strategy three is paper. It is known that its integer order model does not have a stable equilibrium. Let x,y,1-x-y be the fractions of rock, scissors, paper adopters hence the fractional order replicator equation for this game are:

$$D^\alpha x = x(x + 2y - 1), D^\alpha y = y(1 - 2x - y) \quad (4)$$

It is direct to see that there is an internal solution (1/3,1/3,1/3) and that it is locally asymptotically stable (3) if

$$\tan^{-1}(2) > \alpha\pi/2, .7 > \alpha > 0. \quad (5)$$

Hence we have:

Proposition (2): The internal solution (1/3,1/3,1/3) for the fractional order Rock-Scissors-Paper game (4) is locally asymptotically stable if (5) is satisfied.

Some asymmetric games (where the set of strategies and payoff matrix of some players $\underline{x}$ differ from those of another set of players $\underline{y}$ [4] e.g. if the players are males and females) are another example where the dynamics of the equilibrium solutions of the fractional order system significantly differs from its integer order counterpart. Consider the asymmetric game with the payoff matrices:

$$\begin{bmatrix} a & 0 \\ 0 & b \end{bmatrix}, \begin{bmatrix} 0 & c \\ d & 0 \end{bmatrix}, \quad (6)$$

Where a,b,c,d are positive constants. Then the fractional order replicator equations for this game become

$$D^\alpha x = x(1-x)[(a+b)y - b], D^\alpha y = y(1-y)[-(c+d)x + d], \quad (7)$$

Where x,y are the fraction of adopters of the first strategy in both populations. This system admits an internal equilibrium solution

x= d/(c+d), y=b/(a+b)  (8)

Direct stability analysis shows that the characteristic equation of (7) at the internal equilibrium solution (8) is

$$\lambda^2 + abcd/[(a+b)(c+d)] = 0 \quad (9)$$

i.e. both roots are purely imaginary. Using (3) one gets that the internal equilibrium solution is locally asymptotically stable if $1 > \alpha > 0$.

3. Conclusions:

Complex adaptive systems (CAS) are abundant in nature. Almost all biological, economic and social systems are CAS. Game theory is a useful tool in studying such systems. We have shown that fractional order systems are suitable to model CAS. Therefore fractional order game dynamics should be studied. This work is a first step in this direction. It is shown that in some cases e.g. rock-scissors-paper and asymmetric games the results for the fractional order case differs significantly from its integer order counterpart.

Acknowledgement: We thank the referee for his comments.